\documentstyle[12pt]{article}

\begin{document}

\begin{titlepage}
\setcounter{page}{1}
\title{\bf Nonlinear Phase Modification of the Schr\"{o}dinger Equation}
\author{Waldemar Puszkarz\thanks{Electronic address: puszkarz@cosm.sc.edu}
\\
\small{\it  Department of Physics and Astronomy,}
\\
\small{\it University of South Carolina,}
\\
\small{\it Columbia, SC 29208}}
\date{\small (May 15, 1999)}
\maketitle
\begin{abstract}
A nonlinear modification of the  Schr\"{o}dinger equation is proposed in which 
the Lagrangian density for the  Schr\"{o}dinger equation is extended 
by terms polynomial in $\Delta^{m}\ln \left(\Psi^{*}/{\Psi}\right)$ multiplied 
by $\Psi^{*}{\Psi}$. This introduces a homogeneous nonlinearity in a Galilean 
invariant manner through the phase $S$ rather than the amplitude $R$ of the 
wave function $\Psi =R\exp \left( iS\right)$. From this general scheme we 
choose the simplest minimal model defined in some reasonable way. The model 
in question offers the simplest way to modify the Bohm formulation of quantum 
mechanics so as to allow a leading phase contribution to the quantum potential 
and a leading quantum contribution to the probability current removing 
asymmetries present in Bohm's original formulation. It preserves most of 
physically relevant properties of the Schr\"{o}dinger equation including 
stationary states of quantum-mechanical systems. It can be thought of as the 
simplest model of nonlinear quantum mechanics of extended objects among other 
such models that also emerge within the general scheme proposed. The extensions 
of this model to $n$ particles and the question of separability of compound 
systems are studied. It is noted that there exists a weakly separable extension 
in addition to a strongly separable one. The place of the general modification 
scheme in a broader spectrum of nonlinear modifications of the Schr\"{o}dinger 
equation is discussed. It is pointed out that the models it gives rise to have 
a unique definition of energy in that the field-theoretical energy functional 
coincides with the quantum-mechanical one. It is found that the Lagrangian for 
its simplest variant represents the Lagrangian for a restricted version of the 
Doebner-Goldin modification of this equation. It is also noted that a large 
class of particular models generated by this scheme contradict the thesis that 
the homogeneity of a nonlinear Schr\"{o}dinger equation automatically entails 
its weak separability. 

\vskip 0.5cm
\noindent
\end{abstract}
\end{titlepage}

\section{Introduction}


Several years ago Staruszkiewicz \cite{Sta1} \thinspace put forward new, and
what at that time seemed to be unique\footnote{%
As shown in \cite{Pusz1}, the original Staruszkiewicz modification can be
extended in a manner that preserves its characteristic features.} a way of
modifying the Schr\"{o}dinger equation by adding to its action density a
term $a(\Delta S)^{2}$, where $S$ is the quantum-mechanical phase,
historically motivated by his theory of free electromagnetic phase \cite
{Sta2}. It is a pecularity of this modification that the constant $a$ is
dimensionless in natural units, however its dimensionless character is
restricted to the three-dimensional space. In general, in the same system of
units, $a$ has the dimensions of ${\rm meter}^{3-d}$, where $d$ is the
dimension of space in which to construct a theory. One observes that $%
a(\Delta S)^{2}=-a\left[ \Delta \ln \left( \Psi ^{*}/\Psi \right) \right]
^{2}/4$, where\footnote{%
It should be noted that here $S$ is dimensionless for it represents the
angle. This differs from a more common convention in which $\Psi =R\exp
\left( iS/\hbar \right) $ so that $S$ has the dimensions of action. We will
adhere to this convention throughout the most of this paper.} $\Psi =R\exp
\left( iS\right) $. It is this density that together with the Lagrangian
density for the Schr\"{o}dinger equation, 
\begin{equation}
L_{SE}({\vec{r}},t)=\frac{i\hbar }{2}\left( \Psi ^{*}\frac{\partial \Psi }{%
\partial t}-\frac{\partial \Psi ^{*}}{\partial t}\Psi \right) -\frac{{\hbar }%
^{2}}{2m}\vec{\nabla}\Psi ^{*}\vec{\nabla}\Psi -\Psi ^{*}V\Psi ,  \label{1}
\end{equation}
constitutes the integrand of the complete action for the Schr\"{o}dinger
equation with a potential $V$ in the modification proposed by Staruszkiewicz.

The purpose of the present paper is to suggest yet another possible way of
modifying the Schr\"{o}dinger equation which seems to have some advantages
of greater universality and generality compared to Staruszkiewicz's version
and rather insignificant shortcomings. One can view our proposal as a
variation of sorts on his modification which preserves its main theme in
that it grants the phase an important dynamical role in the modified
equations of motion and also in that it employs the nonlinearity of the $%
\Delta S$ term. Nevertheless, it is a completely distinct construction.
Recently, there has been a considerable interest in nonlinear modifications
of quantum mechanics spurned largely by Weinberg's proposal \cite{Wein1,
Wein2} of a relatively general framework for nonlinear quantum mechanics.
However, in spite of this, with the exception of the Staruszkiewicz
modification that precedes Weinberg's, we have not found in the literature a
construction similar to ours.

This paper is organized as follows. We introduce our modification in the
next section where we formulate its general scheme, discuss distinct classes
of particular modifications that emerge within this scheme, and single out
two examples of the simplest nonlinear Schr\"{o}dinger equations
characteristic of these classes. These two models are discussed in more
detail throughout the rest of the paper. In the subsequent section, we
present the simplest minimal version of the modification. In section 4, as
one more way of demonstrating its peculiar properties, we confront the
modification with some notable nonlinear generalizations of the
Schr\"{o}dinger equation put forward earlier. This section is followed by
conclusions where also the motivations for studying the modification
proposed and its conceivable applications are elaborated on.

\section{The Modification}

The basic idea of our modification is to supplement the Lagrangian density
of the Schr\"{o}dinger equation by terms involving the phase in explicit yet
a Galilean-invariant manner by means of the $\ln \left( \Psi ^{*}/\Psi
\right) $ contribution. For this reason, the terms containing $\vec{\nabla}%
\ln \left( \Psi ^{*}/\Psi \right) $ have to be excluded as breaking this
invariance. Indeed, since under the Galilean transformation of coordinates, $%
\vec{x}=\vec{x}^{\prime }+\vec{v}t$, $t^{\prime }=t$, the phase changes as 
\begin{equation}
S^{\prime }=S-m\vec{v}\cdot \vec{x}+\frac{1}{2}m\vec{v}^{2}t,  \label{2}
\end{equation}
the lowest order operator that complies with the condition of Galilean
invariance is $\Delta $. We also require that the modification be
independent of the dimensionality of space in which to develop it. For this
to be accomplished one needs to multiply the $\left[ \Delta ^{k}\ln \left(
\Psi ^{*}/\Psi \right) \right] ^{n}$ terms by $R^{2}={\Psi ^{*}}{\Psi }$, $n$
and $k$ being positive integers. Therefore, the modified Lagrangian density
we propose consists, in addition to (1), of terms 
\begin{equation}
L_{kn}=c_{kn}\Psi ^{*}\left[ \Delta ^{k}\ln \left( \Psi ^{*}/\Psi \right)
\right] ^{n}\Psi =C_{kn}\left( \Delta ^{k}S\right) ^{n}R^{2},  \label{3}
\end{equation}
where $c_{kn}$ and $C_{kn}$ are certain presumably small dimensional
constants, complex (in general) and real, respectively.

Now, the most general Lagrangian density for the modified Schr\"{o}dinger
equation is $L(R,S)=L_{SE}(R,S)+L_{NL}(R,S)$, where, after (1), 
\begin{equation}
-L_{SE}(R,S)=\hbar R^{2}\frac{\partial S}{\partial t}+\frac{\hbar ^{2}}{2m}%
\left[ \left( \vec{\nabla}R\right) ^{2}+R^{2}\left( \vec{\nabla}S\right)
^{2}\right] +R^{2}V,  \label{4}
\end{equation}
and 
\begin{equation}
L_{NL}(R,S)=-\sum_{k,n}C_{kn}\left( \Delta ^{k}S\right) ^{n}R^{2}.  \label{5}
\end{equation}
In principle, $k$ and $n$ run from $1$ to infinity, but for practical
purposes such a structure cannot be deemed very attractive and one needs
either to terminate this series at some point or to choose only some terms
out of it. Before we proceed to the modified Schr\"{o}dinger equation, for
the sake of further discussion, let us first write down the linear
Schr\"{o}dinger equation in the nonlinear Madelung representation \cite{Mad}%
, 
\begin{equation}
\hbar \frac{\partial R^{2}}{\partial t}+\frac{\hbar ^{2}}{m}\vec{\nabla}%
\cdot \left( R^{2}\vec{\nabla}S\right) =0,  \label{6}
\end{equation}
\begin{equation}
\frac{\hbar ^{2}}{m}\Delta R-2R\hbar \frac{\partial S}{\partial t}-2RV-\frac{%
\hbar ^{2}}{m}R\left( \vec{\nabla}S\right) ^{2}=0.  \label{7}
\end{equation}
The modified Schr\"{o}dinger equation in the same formulation is equivalent
to a system of two nonlinear equations that derive from $L(R,S)$, 
\begin{equation}
\hbar \frac{\partial R^{2}}{\partial t}+\frac{\hbar ^{2}}{m}\vec{\nabla}%
\cdot \left( R^{2}\vec{\nabla}S\right) -\sum_{k,n}nC_{kn}\Delta ^{k}\left[
\left( \Delta ^{k}S\right) ^{n-1}R^{2}\right] =0,  \label{8}
\end{equation}
\begin{equation}
\frac{\hbar ^{2}}{m}\Delta R-2R\hbar \frac{\partial S}{\partial t}-2RV-\frac{%
\hbar ^{2}}{m}R\left( \vec{\nabla}S\right) ^{2}-2\sum_{k,n}C_{kn}\left(
\Delta ^{k}S\right) ^{n}R=0,  \label{9}
\end{equation}
where the first of these equations is the continuity equation for the
probability density $\rho =R^{2}$ and the current 
\begin{equation}
\vec{j}=\frac{\hbar ^{2}}{m}R^{2}\vec{\nabla}S-\sum_{k,n}nC_{kn}\vec{\nabla}%
\left[ \Delta ^{k-1}\left( \left( \Delta ^{k}S\right) ^{n-1}R^{2}\right)
\right]  \label{10}
\end{equation}
Due to the higher order derivatives occuring in the Lagrangian densities (3)
and (5), one obtains the above equations from the principle of least action
assuming that not only variations of $S$ and $R$ vanish on the boundary in
the spatial infinity but also variations of their derivatives, such as, for
instance, $\vec{\nabla}S$ and $\vec{\nabla}R$. The energy functionals
attributed to these Lagrangian densities are given by the spatial integrals
of these densities and since they are homogeneous in $R^{2}$ one expects the
total ``modified'' energy to be finite for the configurations that satisfy
the normalization condition $\int d^{3}xR^{2}=1$. For the very same reason
one can weaken our condition on the variations of derivatives by admitting
arbitrary such variations. The boundary terms will then not contribute to
the variations of the total action for their integrals vanish in the spatial
infinity as $R$ vanishes there. It should be noted that the total energy
functional, 
\begin{equation}
E=\int d^{3}x\left\{ \frac{\hbar ^{2}}{2m}\left[ \left( \vec{\nabla}R\right)
^{2}+R^{2}\left( \vec{\nabla}S\right) ^{2}\right] +\sum_{k,n}C_{kn}\left(
\Delta ^{k}S\right) ^{n}R^{2}+VR^{2}\right\} ,  \label{11}
\end{equation}
which derives within the Lagrangian field-theoretical framework as a
constant of motion for configurations that do not depend explicitly on time
(as, for instance, when $V\neq V(t)$) coincides with the quantum-mechanical
one that represents the expectation value of the Hamiltonian operator for
the modification in question \cite{Pusz4}. Since such a property is
relatively rare among the nonlinear modifications of the Schr\"{o}dinger
equation \cite{Pusz4}, it certainly adds to the uniqueness of this class of
modifications and the consistency of their formulation in a manner similar
to the linear Schr\"{o}dinger equation which shares with them the property
in question. One can formulate the modified equation in terms of the entire
wave function $\Psi $ and its complex conjugate $\Psi ^{*}$. We will do this
in the next section for a special version of the modification proposed.

Let us now comment on some particular cases of the modifications that derive
from $L_{kn}$ of (3) as a function of $k$ and $n$. Let us note that the
dimensions of coupling constants associated with $L_{kn}$ and $L_{nk}$ are
the same. However, in general, $L_{kn}$ and $L_{nk}$ do not lead to the same
type of modifications. It is so only when $k\geq 1$ and $n>1$ in which case
the modifications that stem from these Lagrangians describe time-reversible
systems. As seen from equations (6) and (7), for the free linear
Schr\"{o}dinger equation to be invariant under the time-reversal
transformation, $S$ must change the sign. This change of sign is not the
symmetry of a modification if $n=1$ independently of the value of $k$. Such
modifications describe irreversibility in quantum systems. The simplest
example of a nonlinear part of the Lagrangian for these type of
modifications is provided by $L_{11}$. We will denote the total Lagrangian
for this modification, i.e., including also its linear part, by $L_{1}$. The
simplest examplification of a nonlinear part of the Lagrangian for the
former type of modifications is given by $L_{12}$. Let us denote the total
Lagrangian for this modification by $L_{2}$. In what follows, we will
concentrate on these Lagrangians as representatives of the discussed classes
of modifications. To conlude this part, mixing $L_{kl}$ and $L_{1n}\,$for $k$
$\geq n$ and an arbitrary $l$ will lead to equations of motion that are
time-irreversible. It is also for this reason that the general equations
(8-9), without imposing any constraints, describe irreversible systems.

One can approach the problem of irreversibility from completely opposite
angles. One can accept it arguing that it is only due to the smallness of
the coupling constant that the violation of time-reversal has not been
observed yet. We note that one obtains in this way a simple mechanism to
generate the time asymmetry on the quantum level. Or, one can consider the
time-reversal violation a blemish or an unnecessary ingredient of theory in
which case terms that cause it should not be allowed in the construction of
a modified Lagrangian density. We will call a version that does not admit
these terms a minimal phase extension{\it \ }of the Schr\"{o}dinger equation
for it entails the smallest departure from the properties of the free linear
Schr\"{o}dinger equation.

The linear Schr\"{o}dinger equation involves only the mass of a quantum
system $m$ and the Planck constant. Unless some other dimensional parameters
are present one cannot define the characteristic energy scale of a system or
theory. These parameters could be related to the coupling constants of
potentials (or their concomitants, e.g., the frequency $\omega $ in the
potential of harmonic oscillator) or their range, as, for example, is the
case for an infinitely deep potential well of some width $d$. The
modification under discussion does contain such a parameter, the coupling
constant $C_{kn}$. The dimensional analysis shows that in a theory defined
by $m$, $\hbar $, and $C_{kn}$ the energy scales as $E_{kn}=$ $\left( \hbar
^{2k\cdot n}/\left| C_{kn}\right| m^{k\cdot n}\right) ^{1/(k\cdot n-1)}$ ($%
n\cdot k\neq 1$) which in the limit corresponding to linear quantum
mechanics ($C_{kn}\rightarrow 0$) produces infinity! Let us note though that
the characteristic length of such a theory is given by $l_{kn}^{2}=\left(
\left| C_{kn}\right| m/\hbar ^{2}\right) ^{1/(k\cdot n-1)}$. Consequently, 
\begin{equation}
\left| C_{kn}\right| =\frac{\hbar ^{2}l_{kn}^{2(k\cdot n-1)}}{m}=\frac{%
q^{(k\cdot n-1)}\hbar ^{2}\lambda _{c}^{2(k\cdot n-1)}}{m}=\frac{q^{(k\cdot
n-1)}\hbar ^{2k\cdot n}}{m^{2k\cdot n-1}c^{2\left( k\cdot n-1\right) }},
\label{12}
\end{equation}
where $\lambda _{c}=\hbar /mc$ is the Compton wavelength and $%
q=l_{kn}^{2}/\lambda _{c}^{2}$ is some dimensionless number that we will
call the Compton quotient. Now, since $E_{kn}=$ $\hbar ^{2}/ml_{kn}^{2}$, we
see that the most natural way to avoid the infinity in question is to assume
that $l_{kn}^{2}$ is proportional to $l_{c}^{2}$ which represents an
intrinsic property of the system and as such, similarly as the mass of a
quantum-mechanical system, is never equal zero. This leads to a novel type
of theory which can be thought of as quantum mechanics of extended objects
of some characteristic size $l_{c}$, the simplest model being furnished by
strings, and which is necessarily nonlinear. The classical limit of this
theory does exist when $\hbar $ tends to zero, but nevertheless on the level
of equations the theory cannot be reduced to linear quantum mechanics,
similarly as the latter does not boil down to any physically meaningful
theory when the mass of a quantum particle becomes zero. These largely
dimensional arguments do not apply to the modification that derives from $%
L_{1}$ for which no fundamental length exists; the dimension of length
cannot be expressed as a function of dimensions of $m$, $\hbar $, and $C_{1}$
in parallel to linear quantum mechanics. The same holds true for the energy.
The simplest of the modifications that can be interpreted in the way
discussed are provided by $L_{12}$ or $L_{12}$ or a combination of these.
Their coupling constants are proportional to $\left| C_{2}\right| =\hbar
^{2}l_{c}^{2}/m$.

From now on, we will focus our attention on the simplest extensions of our
general scheme which, as pointed out above, stem from Lagrangians $L_{1}$
and $L_{2}$. These are the leading terms in the scheme. As in any
field-theoretical construction, also here we settle for the simplest models.
Only if compelling reasons arise to experiment with higher order terms one
finds doing so justifiable. Hence, our selection at this point is based
mainly on the principle of simplicity which, even if proved extremely useful
in physics, belongs to the realm of aesthetics or methodology rather than
physics proper. Other, physical, principles that one would like to use in
order to discriminate between these two models or argue for their uniqueness
have either already been invoked or still will. It is also the principle of
simplicity that ensures the smallest departure of the discussed models from
the linear Schr\"{o}dinger equation. In what follows, the general scheme we
have introduced will be used only in a few circumstances in order to make
some more general observations.

We discuss the modification that derives from $L_{2}$ in the next section.
Since $L_{1}$ gives rise to the modification that constitutes a part of a
more general scheme already proposed by Doebner and Goldin \cite{Doeb1,
Doeb2}, we elaborate on this modification in the context of Doebner-Goldin
proposal in section 4.

\section{The Simplest Minimal Phase Extension (SMPE)}

The simplest of minimal phase extensions stems from $L_{2}$ introduced in
the previous section. It involves the minimal number of derivatives of the
lowest possible order while preserving the basic features of the free linear
Schr\"{o}dinger equation such as, apart from the Galilean invariance, the
invariance under the space and time reflections. Other features of this
equation may however be compromised. This, as we will see, is the case for
the weak separability of composite systems \cite{Bial}. The total Lagrangian
for this modification is 
\begin{equation}
-L_{SMPE}(R,S)=\hbar R^{2}\frac{\partial S}{\partial t}+\frac{\hbar ^{2}}{2m}%
\left[ \left( \vec{\nabla}R\right) ^{2}+R^{2}\left( \vec{\nabla}S\right)
^{2}\right] +CR^{2}\left( \Delta S\right) ^{2}+R^{2}V,  \label{13}
\end{equation}
where as argued before one can identify the coupling constant $C$ with $%
\hbar ^{2}l_{c}^{2}/m$, $l_{c}$ being the characteristic size of an extended
particle-system. The complete energy functional for the SMPE derived from
this Lagrangian, 
\begin{equation}
E=\int \,d^{3}x\,\left\{ \frac{\hbar ^{2}}{2m}\left[ \left( \vec{\nabla}%
R\right) ^{2}+R^{2}\left( \vec{\nabla}S\right) ^{2}\right] +CR^{2}\left(
\Delta S\right) ^{2}+VR^{2}\right\} ,  \label{14}
\end{equation}
clearly exhibits good convergent properties for square integrable wave
functions. The equations of motion that derive from (13) read in the
hydrodynamic formulation 
\begin{equation}
\hbar \frac{\partial R^{2}}{\partial t}+\frac{\hbar ^{2}}{m}\vec{\nabla}%
\cdot \left( R^{2}\vec{\nabla}S\right) -2C\Delta \left( R^{2}\Delta S\right)
=0,  \label{15}
\end{equation}
\begin{equation}
\frac{\hbar ^{2}}{m}\Delta R-2R\hbar \frac{\partial S}{\partial t}-2RV-\frac{%
\hbar ^{2}}{m}R\left( \vec{\nabla}S\right) ^{2}-2CR\left( \Delta S\right)
^{2}=0.  \label{16}
\end{equation}
One observes that any stationary solution to the linear Schr\"{o}dinger
equation is also a solution to these equations. Indeed, since for such
solutions $S=-Et/\hbar +const$ , the above equations reduce in this case to
the Schr\"{o}dinger-Madelung equations. There may however exist other
stationary solutions to (15-16); they are supposed to satisfy the condition $%
\partial R^{2}/\partial t=0$. It is only for the Schr\"{o}dinger equation
that this condition ensures that the phase is a unique linear function of
time only. Unlike the discussed version of the modification, its variant
generated by $L_{1}$ does not allow the stationary solutions for which $%
S=-Et/\hbar $ $+const$ and therefore affects rather dramatically the
stationary states of known physical systems such as the hydrogen atom or
harmonic oscillator.

One can also put the modified Schr\"{o}dinger equation in terms of $\Psi $
and $\Psi ^{*}$. To this end one expresses $L_{SMPE}$ in terms of these
variables and derives from it the Euler-Lagrange equation for $\Psi ^{*}$.
The Lagrangian in question reads 
\begin{equation}
L_{SMPE}(\Psi ,\Psi ^{*})=\frac{i\hbar }{2}\left( \Psi ^{*}\frac{\partial
\Psi }{\partial t}-\frac{\partial \Psi ^{*}}{\partial t}\Psi \right) -\frac{{%
\hbar }^{2}}{2m}\vec{\nabla}\Psi ^{*}\vec{\nabla}\Psi -\Psi ^{*}V\Psi +\frac{%
C}{4}P^{2}\Psi \Psi ^{*},  \label{17}
\end{equation}
where the factor $C/4$ was chosen so as to reproduce the equations of motion
in the hydrodynamic form of (15-16). The result of the derivation turns out
to be 
\begin{equation}
i\hbar \frac{\partial \Psi }{\partial t}=H_{SMPE}\Psi =\left( -\frac{\hbar
^{2}}{2m}\Delta +V\right) \Psi -\frac{C}{4}G[\Psi ,\Psi ^{*}]\Psi ,
\label{18}
\end{equation}
where 
\begin{equation}
G[\Psi ,\Psi ^{*}]=P^{2}+\frac{2\Delta \left( \Psi \Psi ^{*}P\right) }{\Psi
\Psi ^{*}}  \label{19}
\end{equation}
and 
\begin{equation}
P=\Delta \ln (\frac{\Psi ^{*}}{\Psi }).  \label{20}
\end{equation}

The classical limit of this modification in the sense of the Ehrenfest
theorem may not always exist since the standard Ehrenfest theorem of linear
quantum mechanics is altered by nonlinear corrections. Let us now work out
these corrections. For a general observable $A$ one finds that 
\begin{equation}
\frac{d}{dt}\left\langle A\right\rangle =\frac{d}{dt}\left\langle
A\right\rangle _{L}+\frac{d}{dt}\left\langle A\right\rangle _{NL},
\label{21}
\end{equation}
where the nonlinear contribution due to $H_{NL}[\Psi ,\Psi
^{*}]=H_{R}+iH_{I} $ can be expressed as 
\begin{equation}
\frac{d}{dt}\left\langle A\right\rangle _{NL}=\left\langle \left\{
A,H_{I}\right\} \right\rangle -i\left\langle \left[ A,H_{R}\right]
\right\rangle ,  \label{22}
\end{equation}
with $H_{R}$ and $H_{I}$ being the real and imaginary part of $H_{NL}=-CG/4$%
, correspondingly. The brackets $<>$ denote the mean value of the quantity
embraced, $[\cdot ,\cdot ]$ and $\{\cdot ,\cdot \}$ denote commutators and
anticommutators, respectively. Specifying $A$ for the position and momentum
operators, one obtains the general form of the modified Ehrenfest relations 
\begin{equation}
m\frac{d}{dt}\left\langle \vec{r}\right\rangle =\left\langle \vec{p}%
\right\rangle +I_{1},  \label{23}
\end{equation}
\begin{equation}
\frac{d}{dt}\left\langle \vec{p}\right\rangle =-\left\langle \vec{\nabla}%
V\right\rangle +I_{2},  \label{24}
\end{equation}
where 
\begin{equation}
I_{1}=\frac{2m}{\hbar }\int \,d^{3}x\vec{r}R^{2}H_{I},  \label{25}
\end{equation}
\begin{equation}
I_{2}=\int \,d^{3}xR^{2}\left( 2H_{I}\vec{\nabla}S-\vec{\nabla}H_{R}\right)
-i\int d^{3}x\vec{\nabla}\left( R^{2}H_{I}\right) .  \label{26}
\end{equation}
The imaginary term in the last formula can be discarded for homogeneous
modifications such as the one in question. For this type of modifications, $%
H_{I}=f(R^{2})/R^{2}$, where $f(R^{2})$ is a certain operator acting on $%
R^{2}$ that can also involve the phase. Now, for square integrable wave
functions for which $R$ vanishes sufficiently fast in the infinity, we have $%
\int d^{3}x\vec{\nabla}\left( R^{2}H_{I}\right) =\int d^{2}x\vec{n}%
f(R^{2})=0 $, where $\vec{n}$ is the unit vector normal to the boundary in
the infinity. In the case under study, $H_{R}=C\left( \Delta S\right) ^{2}$
and $H_{I}=C\Delta \left( \Delta SR^{2}\right) /R^{2}$, therefore the
Ehrenfest relations for the SMPE read 
\begin{equation}
m\frac{d}{dt}\left\langle \vec{r}\right\rangle =\left\langle \vec{p}%
\right\rangle +\frac{2Cm}{\hbar }\int \,d^{3}x\vec{r}\Delta \left( \Delta
SR^{2}\right) ,  \label{27}
\end{equation}
\begin{equation}
\frac{d}{dt}\left\langle \vec{p}\right\rangle =-\left\langle \vec{\nabla}%
V\right\rangle +C\int \,d^{3}x\left[ 2\vec{\nabla}S\Delta \left( \Delta
SR^{2}\right) -R^{2}\vec{\nabla}\left( \Delta S\right) ^{2}\right] .
\label{28}
\end{equation}
The Ehrenfest relations are Galilean invariant as are their nonlinear
contributions for the wave functions that satisfy the equations of motion.

We see that, in general, the nonlinear corrections to the Ehrenfest
relations do not vanish, leading to a different classical limit than in the
linear theory. This feature of the modification is shared by other nonlinear
generalizations of the Schr\"{o}dinger equation, the only notable exception
from this rule being the Bia\l ynicki-Birula and Mycielski modification, for
which $H_{I}=0$ and $\vec{\nabla}H_{R}=\vec{\nabla}R^{2}/R^{2}$ so that both 
$I_{1}$ and $I_{2}$ vanish, the first identically and the second one for
normalizable wave functions. Some of these modifications do possess the
Ehrenfest limit, but only for certain values of their parameters. This, for
instance, applies to the Doebner-Goldin modification. That the modification
discussed does not have such a limit suggests that its equations are not
linearizable, i.e., they cannot be transformed into the form of the linear
Schr\"{o}dinger equation, and thus are likely to contain some new physics
that cannot be described by linear theory. This observation is corraborated
by the fact that the Doebner-Goldin equations in their Ehrenfest domain are
linearizable. It is probably justified to expect that a linearizable
modification possesses the Ehrenfest limit. The converse may not be true,
that is, a modification that has the classical limit does not have to be
linearizable as seems to be the case for the Bia\l ynicki-Birula and
Mycielski modification. The modified Ehrenfest relations may still allow for
the classical limit for some special wave functions for which both $I_{1}$
and $I_{2}$ vanish. On the other hand, perhaps due to its simplicity, the
relevance of the Ehrenfest theorem tends to be overestimated. As shown in 
\cite{Ball}, this theorem is neither sufficient nor necessary to
characterize the classical regime of quantum theory. It is easy to convince
oneself that in one dimension as long as $\Delta S$ is an arbitrary function
of time the Ehrenfest theorem holds also for the modification under study.
The same holds true in higher dimensions for factorizable wave functions.

The modification of ours when cast in the Bohmian framework of quantum
mechanics offers interesting and rather a logical extension of it. Let us
recall that in Bohm's approach to quantum mechanics a part of the linear
Schr\"{o}dinger equation (7) can be put as\footnote{%
To arrive at this form of quantum potential, one needs to restore $\hbar $
in the phase of the wave function so that $\Psi =R\exp (i{\cal S}/\hbar )$,
where now ${\cal S\ }$has the dimensions of action. This is formally
equivalent to changing $S\rightarrow {\cal S}/\hbar $.} 
\begin{equation}
\frac{\partial {\cal S}}{\partial t}+\frac{1}{2m}\left( \vec{\nabla}{\cal S}%
\right) ^{2}+V_{C}+V_{Q}^{Sch}=0,  \label{29}
\end{equation}
where $V_{Q}^{Sch}=V_{R}=-\hbar ^{2}\Delta R/2mR$ represents the quantum
potential of Bohm for the Schr\"{o}dinger equation \cite{Bohm}. Without the
external ``classical'' potential $V_{C}$ this equation reduces to the
ordinary Hamilton-Jacobi equation for the phase ${\cal S}$ in the quantum
potential $V_{Q}^{Sch}$. It is somewhat puzzling that this potential does
not depend on the phase itself, as in general one would expect the evolution
of a quantum system driven by the very potential to depend not only on the
amplitude of its wave function but the phase as well. In the proposed
modification the quantum potential $V_{Q}^{Sch}$ is supplemented by a term
that involves also the phase, which eliminates the asymmetry in question. As
a result of this contribution, $V_{{\cal S}}=C\left( \Delta {\cal S}\right)
^{2}/2m\hbar ^{2}$, the quantum potential becomes $V_{Q}=V_{R}+V_{{\cal S}}$%
. One can interpret this additional term as a backreaction term, due to the
reaction of the phase on the evolution in the potential $V_{R}$. Because of
the presence of $\hbar ^{2}$ in the denominator of $V_{{\cal S}}$, this
component of the quantum potential would dominate the other term of $V_{Q}$,
were it not, as argued earlier, for the smallness of $C=q_{\pm }\hbar
^{4}/m^{3}c^{2}$, where $q_{\pm }$ is a real number whose absolute value
equals the Compton quotient but whose sign is undetermined by the theory as
signalled by its subscript. This enables one to write the quantum potential
in a more succinct way, 
\begin{equation}
V_{Q}=\hbar ^{2}\left[ \frac{q_{\pm }}{2m^{3}c^{2}}\left( \Delta {\cal S}%
\right) ^{2}-\frac{1}{2m}\frac{\Delta R}{R}\right] .  \label{30}
\end{equation}
In the Bohmian framework of linear Schr\"{o}dinger equation the continuity
equation (6) looses any trace of the Planck constant, which reveals an even
greater asymmetry of this approach. As opposed to equation (7) that contains
``quantum'' contributions, i.e., terms proportional to $\hbar ^{2}$,
equation (6) looks like a classical equation for a ``classical'' current $%
\vec{j}_{C}=R^{2}\vec{\nabla}{\cal S}/m$. This situation is ``corrected'' in
the proposed modification where the probability current contains a quantum
contribution as well, 
\begin{equation}
\vec{j}_{Q}=\frac{2q_{\pm }\hbar ^{2}}{m^{3}c^{2}}\vec{\nabla}\left(
R^{2}\Delta {\cal S}\right) ,  \label{31}
\end{equation}
and the total current reads 
\begin{equation}
\vec{j}=\frac{1}{m}R^{2}\vec{\nabla}{\cal S}+\frac{2q_{\pm }\hbar ^{2}}{%
m^{3}c^{2}}\vec{\nabla}\left( R^{2}\Delta {\cal S}\right) .  \label{32}
\end{equation}
One can now write equations (15-16) in a very compact and symmetrical way, 
\begin{equation}
\frac{\partial R^{2}}{\partial t}+\vec{\nabla}\cdot \left( \vec{j}_{C}+\vec{j%
}_{Q}\right) =0,  \label{33}
\end{equation}
\begin{equation}
\frac{\partial S}{\partial t}+\frac{1}{2m}\left( \vec{\nabla}{\cal S}\right)
^{2}+V_{C}+V_{Q}=0,  \label{34}
\end{equation}
where both $\vec{j}_{Q}$ and $V_{Q}$ contain terms proportional to $\hbar
^{2}$. The SMPE offers the simplest way to extend the Bohm approach to
quantum mechanics by incorporating a quantum phase contribution to the
quantum potential and a quantum contribution to the probability current. Let
us note that these new terms are by no means negligible being of the same
order of magnitude in terms of $\hbar $ and $c$ as the spin-orbit coupling
responsible for the fine structure of atomic spectra. It is therefore
conceivable even if speculative that the discussed terms describe some
physically relevant ``hydrodynamic'' fine structure of the quantum world
that owing to the elusive nature of the phase has somehow managed to escape
our attention.

Let us note that the coupling constant $C$ has the same dimensions as the
coupling constant that emerges in the leading relativistic approximation to
the Schr\"{o}dinger equation. One obtains this approximation from the
relativistic relation between the energy $E$ and the momentum $p$ of a
single particle of mass $m$ truncated to 
\begin{equation}
E=\frac{p^{2}}{2m}-\frac{p^{4}}{8m^{3}c^{2}},  \label{35}
\end{equation}
which upon the first quantization leads to the modified Schr\"{o}dinger
equation, 
\begin{equation}
i\hbar \frac{\partial \Psi }{\partial t}=\left( -\frac{\hbar ^{2}}{2m}\Delta
-\frac{\hbar ^{4}}{8m^{3}c^{2}}\Delta ^{2}\right) \Psi .  \label{36}
\end{equation}
Despite the same dimensions of these coupling constants, the last equation
does not coincide with our modification in the hydrodynamic representation,
even in the Galilean limit.

The continuity equation for this modification can also be cast in the form
of the generalized Fokker-Planck equation 
\begin{equation}
\frac{\partial W}{\partial t}+\frac{\partial }{\partial x^{i}}(D_{i}W+\frac{%
\partial }{\partial x^{j}}D_{ij}W)=0,  \label{37}
\end{equation}
where $D_{i}$ and $D_{ij}$ are some vector and tensor object, respectively,
and the summation over repeated indices is assumed. This is not unlike in
the other particular version of our general scheme that stems from $L_{1}$,
originally proposed by Doebner and Goldin \cite{Doeb1} by adopting the
simplest form of continuity equation of the Fokker-Planck type. One easily
identifies $W$ with $R^{2}$, $mD_{i}$ with $\partial {\cal S}/\partial x^{i}$%
, and $D_{ij}$ with $-2C\delta _{ij}\Delta {\cal S}$. On the other hand, the
energy equation (16) when written in the form 
\begin{equation}
\frac{\partial {\cal S}}{\partial t}+\frac{1}{2m}\left( \vec{\nabla}{\cal S}%
\right) ^{2}+B\left( \Delta {\cal S}\right) ^{2}-\frac{\hbar ^{2}}{2m}\frac{%
\Delta R}{R}+V=0,  \label{38}
\end{equation}
can be thought of as a generalization of the Hamilton-Jacobi equation with $%
B=C/\hbar ^{2}$.

It is a general feature of nonlinear wave mechanics that in a system
consisting of two particles the very existence of one of them affects the
wave function of the other one. Thus, even in the absence of forces the rest
of the world influences the behavior of an isolated particle. As emphasized
in \cite{Bial}, this is not necessarily a physically sound situation. A way
to avoid it that have come to be known as the weak separability \cite{Czach1}
was suggested by Bia{\l }ynicki-Birula and Mycielski \cite{Bial}. As we will
demonstrate it, in its most straightforward multi-particle extension, the
SMPE does not permit the weak separability of composite systems. To explain
why this is so, let us first demonstrate the weak separability of the linear
Schr\"{o}dinger equation in the hydrodynamic formulation.

We are considering a quantum system made up of two noninteracting subsystems
in the sense that 
\begin{equation}
V(x_{1},x_{2},t)=V_{1}(x_{1},t)+V_{2}(x_{2},t).  \label{39}
\end{equation}
We will show that a solution of the Schr\"{o}dinger equation for this system
can be put in the form of the product of wave functions for individual
subsystems for any $t>0$, that is, $\Psi (x_{1},x_{2},t)=\Psi
_{1}(x_{1},t)\Psi _{2}(x_{2},t)=R_{1}(x_{1},t)R_{2}(x_{2},t)exp\left\{
i(S_{1}(x_{1},t)+S_{2}(x_{2},t))\right\} $ and that this form entails the
separability of the subsystems. The essential element here is that the
subsystems are initially uncorrelated which is expressed by the fact that
the total wave function is the product of $\Psi _{1}(\vec{x}_{1},t)$ and $%
\Psi _{2}(\vec{x}_{2},t)$ at $t=0$. The assumption that the compound wave
function is the product one is what defines this form of separability. What
we will show then is that the subsystems remain uncorrelated during the
evolution and that, at the same time, they also remain separated. It is the
additive form of the total potential that guarantees that no interaction
between the subsystems occurs, ensuring that they remain uncorrelated during
the evolution. However, such an interaction may, in principle, occur in
nonlinear modifications of the Schr\"{o}dinger equation even if the form of
the potential itself does not imply that. This is due to a coupling that the
nonlinear term of the equation usually causes between $\Psi _{1}(\vec{x}%
_{1},t)$ and $\Psi _{2}(\vec{x}_{2},t)$.

The Schr\"{o}dinger equation for the total system, assuming for simplicity
that the subsystems have the same mass $m$, reads now

\begin{eqnarray}
\lefteqn{\hbar \frac{\partial R_{1}^{2}R_{2}^{2}}{\partial t}+\frac{\hbar
^{2}}{m}\left\{ \left( \vec{\nabla}_{1}+\vec{\nabla}_{2}\right) \cdot \left[
R_{1}^{2}R_{2}^{2}\left( \vec{\nabla}_{1}S_{1}+\vec{\nabla}_{2}S_{2}\right)
\right] \right\} =\hbar R_{2}^{2}\frac{\partial R_{1}^{2}}{\partial t}+\hbar
R_{1}^{2}\frac{\partial R_{2}^{2}}{\partial t}}  \nonumber \\
&&+\frac{\hbar ^{2}}{m}R_{2}^{2}\vec{\nabla}_{1}\cdot \left( R_{1}^{2}\vec{%
\nabla}_{1}S_{1}\right) +\frac{\hbar ^{2}}{m}R_{1}^{2}\vec{\nabla}_{2}\cdot
\left( R_{2}^{2}\vec{\nabla}_{2}S_{2}\right) =R_{1}^{2}R_{2}^{2}\left\{
\left[ \hbar \frac{1}{R_{1}^{2}}\frac{\partial R_{1}^{2}}{\partial t}%
+\right. \right.  \nonumber \\
&&\left. \left. \frac{\hbar ^{2}}{m}\frac{1}{R_{1}^{2}}\vec{\nabla}_{1}\cdot
\left( R_{1}^{2}\vec{\nabla}_{1}S_{1}\right) \right] +\left[ \hbar \frac{1}{%
R_{2}^{2}}\frac{\partial R_{2}^{2}}{\partial t}++\frac{\hbar ^{2}}{m}\frac{1%
}{R_{2}^{2}}\vec{\nabla}_{2}\cdot \left( R_{2}^{2}\vec{\nabla}%
_{2}S_{2}\right) \right] \right\} =0  \label{40}
\end{eqnarray}

and

\begin{eqnarray}
\lefteqn{\frac{\hbar ^{2}}{m}\left( \Delta _{1}+\Delta _{2}\right)
R_{1}R_{2}-2\hbar R_{1}R_{2}\frac{\partial (S_{1}+S_{2})}{\partial t}-\frac{%
\hbar ^{2}}{m}R_{1}R_{2}\left( \vec{\nabla}_{1}S_{1}+\vec{\nabla}%
_{2}S_{2}\right) ^{2}-}  \nonumber \\
&&\left( V_{1}+V_{2}\right) R_{1}R_{2}=\frac{\hbar ^{2}}{m}R_{2}\Delta
_{1}R_{1}+\frac{\hbar ^{2}}{m}R_{1}\Delta _{2}R_{2}-2\hbar R_{1}R_{2}\frac{%
\partial S_{1}}{\partial t}-2\hbar R_{1}R_{2}\frac{\partial S_{2}}{\partial t%
}  \nonumber \\
&&+\frac{\hbar ^{2}}{m}R_{1}R_{2}\left( \vec{\nabla}_{1}S_{1}\right) ^{2}+%
\frac{\hbar ^{2}}{m}R_{1}R_{2}\left( \vec{\nabla}_{2}S_{2}\right)
^{2}-V_{1}R_{1}R_{2}-V_{2}R_{1}R_{2}=  \nonumber \\
&&R_{1}R_{2}\left\{ \left[ \frac{\hbar ^{2}}{m}\frac{\Delta _{1}R_{1}}{R_{1}}%
-2\hbar \frac{\partial S_{1}}{\partial t}+\frac{\hbar ^{2}}{m}\left( \vec{%
\nabla}_{1}S_{1}\right) ^{2}-V_{1}\right] +\left[ \frac{\hbar ^{2}}{m}\frac{%
\Delta _{2}R_{2}}{R_{2}}-2\hbar \frac{\partial S_{2}}{\partial t}+\right.
\right.  \nonumber \\
&&\left. \left. \frac{\hbar ^{2}}{m}\left( \vec{\nabla}_{2}S_{2}\right)
^{2}-V_{2}\right] \right\} =0.  \label{41}
\end{eqnarray}
Implicit in the derivation of these equations is the fact that $\vec{\nabla}%
_{1}f_{1}\cdot \vec{\nabla}_{2}g_{2}=0$, where $f_{1}$ and $g_{2}$ are
certain scalar functions defined on the configuration space of particle 1
and 2, correspondingly. What we have obtained is a system of two equations,
each consisting of terms (in square brackets) that pertain to only one of
the subsystems. By dividing the first equation by $R_{1}^{2}R_{2}^{2}$ and
the second one by $R_{1}R_{2}$, one completes the separation of the
Schr\"{o}dinger equation for the compound system into the equations for the
subsystems. Morover, we have also showed that indeed the product of wave
functions of the subsystems evolves as the wave function of the total
system. This is, however, not so for the discussed version of our
modification as seen from (15-16). For instance, the second of these
equations contains the term $R\left( \Delta S\right) ^{2}$ which is
nonseparable due to the coupling $\Delta _{1}S_{1}\Delta _{2}S_{2}$. It is
due to similar couplings that other particular modifications that emerge
from our general scheme for $n\ge 2$ also violate the weak separability of
compound systems.

The simplest way to avoid the nonseparability is to change the Lagrangian
(or the equations) for the multi-particle case. The equations for two
particles that we considered above stem from the general $N$-particle
Lagrangian, 
\begin{equation}
-L_{n}(R,S)=\hbar R^{2}\frac{\partial S}{\partial t}+\sum_{i=1}^{N}\frac{%
\hbar ^{2}}{2m_{i}}\left[ \left( \vec{\nabla}_{i}R\right) ^{2}+R^{2}\left( 
\vec{\nabla}_{i}S\right) ^{2}\right] +CR^{2}\left( \sum_{i=1}^{N}\Delta
_{i}S\right) ^{2}+R^{2}V,  \label{42}
\end{equation}
which is nonseparable because of the discussed coupling of terms, $\Delta
_{i}S\Delta _{j}S$. However, if instead of this Lagrangian we use 
\begin{equation}
-L_{n}(R,S)=\hbar R^{2}\frac{\partial S}{\partial t}+\sum_{i=1}^{N}\frac{%
\hbar ^{2}}{2m_{i}}\left[ \left( \vec{\nabla}_{i}R\right) ^{2}+R^{2}\left( 
\vec{\nabla}_{i}S\right) ^{2}\right] +CR^{2}\sum_{i=1}^{N}\left( \Delta
_{i}S\right) ^{2}+R^{2}V,  \label{43}
\end{equation}
we obtain perfectly weakly separable equations. Moreover, the coupling
constant $C$ can be now made particle-dependent, i.e., $C=C_{i}$. The most
natural way to do this is by employing the characteristic size of the
particle $l_{c}$ using the relation $C=\pm \hbar ^{2}l_{c}^{2}/m$. It is in
this theory that the characteristic size of the particle retains its
physical meaning in the case of many particles. Both Lagrangians reduce to
the same expression for one particle, but for many particles they describe
different theories. If one wants to have a theory that does not rule out
separability, one should choose the Lagrangian given by (43). This, however,
does not guarantee that more general compound systems described by
nonfactorizable wave functions will not turn out to be nonseparable. This
weakly separable extension is truly unique in that it does not split the
total wave function into separate one-particle wave functions as would be
the case for the cubic nonlinear equation \cite{Pusz1}. Such an extension is
possible only for homogeneous nonlinear modifications and, in particular, it
can be performed for all the models of the general modification scheme
defined by the Lagrangian (4-5).

On the other hand, as asserted by Czachor \cite{Czach1}, this modification
is separable for any class of quantum-mechanical systems including also
entangled systems in a novel alternative approach of strong separability
(see also \cite{Pol, Jord} where this approach originated). This particular
formalism of strong separability that chooses as its starting point the
nonlinear von Neunmann equation for density matrices \cite{Czach2} admits a
broader family of nonlinearities than the weak separability which we used
for the above demonstration.

\section{Comparison with Other Nonlinear Modifications}

To this day a number of other nonlinear generalizations of the
Schr\"{o}dinger equation have been proposed. In this section we will discuss
some of them stressing similarities and differences with our proposal as yet
another way of presenting its characteristic properties.

To this end, let us start from the Staruszkiewicz modification, the closest
in spirit to the one presented here. In contradistinction to our
generalization, Staruszkiewicz's is nonhomogeneous and, consequently, the
dimensions of its coupling constant depend on the dimensionality of
space-time. It is the presence of the homogeneous $\Psi ^{*}{\Psi }$ terms
that makes the coupling constants in our scheme independent of this
dimensionality, which certainly gives it some greater universality in a
fashion similar to that of the original Schr\"{o}dinger equation. Moreover,
it is due to the same reason that in the Staruszkiewicz modification the
field-theoretical definition of energy as a certain conserved quantity
differs from the quantum-mechanical one that treats energy as the
expectation value of the Hamiltonian operator \cite{Pusz4}. It is also
because of the nonhomogeneity that the Staruszkiewicz modification is not
weakly separable and cannot be made weakly separable in a way that would
comply with the strong separability in the fundamentalist approach \cite
{Pusz1} to this issue.  However, it can be made strongly separable in
Czachor's effective approach \cite{Czach1}.

Arguably, in the recent years, the most studied nonlinear modifications of
the Schr\"{o}dinger equation have been the modification proposed by Bia{\l }%
ynicki-Birula and Mycielski \cite{Bial}\ and that of Doebner and Goldin%
\footnote{%
As shown in \cite{Doeb3}, the modifications in question can be viewed as
belonging to the same family that also includes the Kostin modification \cite
{Kos}, and are related by some generalized nonlinear gauge transformation.} 
\cite{Doeb1, Doeb2}. The basic physical condition that selects the form of
nonlinear terms in each of these modifications is that of weak separability
of composite systems.\footnote{%
As recently demonstrated by L\"{u}cke \cite{Luc1, Luc2}, this by no means
ensures the separability in a more general sense, that is, when a compound
system is described by a nonfactorizable wave function.} It is this property
of the separability that lead Bia{\l }ynicki-Birula and Mycielski to their
model of nonlinearity in which no correlations are introduced by the
nonlinear term. However, this choice would not be unique without another
postulate that greatly limits the class of nonlinear terms. Namely, it is
also stipulated in \cite{Bial} that the only nonlinear terms of the
Lagrangian density for this modification are potential terms that do not
contain any derivatives. By weakening this condition one would end up with a
multitude of admissible terms not unlike in the general scheme of our
modification. Indeed, the possible terms would then include not only the
unique nonlinear potential $V_{mod}=b\ln \left( a|\Psi |^{2}\right) $, where 
$a$ and $b$ are the only undetermined dimensional constants, but also any
term of the form $LV_{mod}$, where $L$ is an arbitrary scalar linear
operator, as for example $\Delta $ or $\Delta ^{2}$. One should note that
this stipulation does not carry any physical contents, except for ensuring
the standard Ehrenfest limit whose physical relevance could be limited even
in the linear theory as pointed out earlier \cite{Ball}. It is just an
additional simplifying assumption without which the unique pick of the
physically desirable nonlinearity is not attainable. As we have demonstrated
it in the preceding section, our modification does possess a similar unique
construction, based on some reasonable postulates, and it involves only one
free parameter. Nevertheless, the fact that the Bia{\l }ynicki-Birula and
Mycielski modification does not alter the standard Ehrenfest relations makes
it exceptional among nonlinear modifications of the Schr\"{o}dinger
equation. Other proposals either admit the Ehrenfest theorem for certain
values of their free parameters only or for wave functions of particular
properties. It has been found that the term $b\ln \left( a|\Psi |^{2}\right) 
$ can be given meaningful physical interpretations. One can see it as the
effect of statistical uncertainty in the form of the potential \cite{Hugh}
or as the potential energy associated with the information encoded or stored
in the distribution of matter described by the probability density $|\Psi
|^{2}$ \cite{Brash}. Although the modification in question was not
originally intended to apply to quantum systems of finite size, as argued in 
\cite{Heft}, if reinterpreted as a theory of extended objects it might be
applicable to the nuclear realm. The modification itself does not imply such
an interpretation unlike the SMPE. One observes that in the modification of
Bia{\l }ynicki-Birula and Mycielski the nonlinear term contains no phase
contribution in which it is complementary to the modification of ours that
introduces nonlinearity only through the phase. What they have in common is
the use of a logarithmic nonlinearity.

The modification derived from $L_{1}$ has the same continuity equation as
the Doebner-Goldin modification. In fact, it constitutes a special case of
this modification. As opposed to the Bia{\l }ynicki-Birula and Mycielski,
the Staruszkiewicz, and the SMPE modifications, in its full-fledged form,
the Doebner-Goldin generalization was intended to describe only a certain
domain of quantum realm, irreversible and dissipative quantum systems. A
strong argument in favor of this model of nonlinearity is lent by group
theoretical analysis of the representations of the {\it Diff}({\bf R}$^{3}$)
group which was proposed as a ``universal quantum kinematical group'' \cite
{Sharp}. In the quantum-mechanical context, particular terms of the general
Doebner-Goldin scheme had appeared well before the fully developed scheme
was put forward. To the best of our knowledge, as a way to modify the
Schr\"{o}dinger equation, a homogeneous term of the type considered by
Doebner and Goldin, was first explicitly employed by Rosen \cite{Ros2} in
1965 but conceived by him even earlier \cite{Ros1}. This term, $\Delta R/R$
in our notation, subsequently appears in the work of other authors guided by
rather diverse motivations. In \cite{Guer, Smo1, Smo2, Vig, Ber}, it appears
solo while in \cite{Kal, Sab, Aub} as part of larger combination. Moreover,
a different term of the general scheme in question, which does not preserve
the Galilean invariance of the modified Schr\"{o}dinger equation, was used
in \cite{Kib}.

One can easily extend the Lagrangian density $L_{1}$ to the density from
which a more general form of the Doebner-Goldin modification emerges, but
which still constitutes a restricted variant of the full-fledged
modification in question. To demonstrate this, let us start from the
Doebner-Goldin modification in the form that is a slight variation on its
original form \cite{Pusz2}, 
\begin{equation}
i\hbar \frac{\partial \Psi }{\partial t}=\left( -\frac{\hbar ^{2}}{2m}\Delta
+V\right) \Psi -\frac{i\hbar D}{2}F_{\left\{ a\right\} }\left[ \Psi ,\Psi
^{*}\right] \Psi +\hbar DF_{\left\{ b\right\} }\left[ \Psi ,\Psi ^{*}\right]
\Psi ,  \label{44}
\end{equation}
where 
\begin{equation}
F_{\left\{ x\right\} }\left[ \Psi ,\Psi ^{*}\right]
=\sum_{i=1}^{n}x_{i}F_{i}\left[ \Psi ,\Psi ^{*}\right]  \label{45}
\end{equation}
and $x_{i}$ are some dimensionless coefficients that form a generic array $%
\left\{ x\right\} $ while $F_{i}\left[ \Psi ,\Psi ^{*}\right] $ are
functionals of $\Psi $ and $\Psi ^{*}$ homogeneous of degree zero in these
functions. The coupling constant $D$ has the dimensions of the diffusion
coefficient, meter$^{2}$second$^{-1}$. In the hydrodynamic representation,
the general form of the functional employed by Doebner and Goldin is \cite
{Pusz2} 
\begin{equation}
F_{\left\{ x\right\} }^{DG}\left[ \rho ,S\right] =x_{1}\Delta S+x_{2}\vec{%
\nabla}S\cdot \left( \frac{\vec{\nabla}\rho }{\rho }\right) +x_{3}\frac{%
\Delta \rho }{\rho }+x_{4}\left( \frac{\vec{\nabla}\rho }{\rho }\right)
^{2}+x_{5}\left( \vec{\nabla}S\right) ^{2},  \label{46}
\end{equation}
where $\rho =R^{2}$. The imaginary part of the Schr\"{o}dinger equation is
supposed to give a continuity equation. The standard way to obtain it is to
multiply both sides of the Schr\"{o}dinger equation by $\Psi ^{*}$ and to
take the imaginary part of the ensuing expression. Now, if $\rho F_{\left\{
a\right\} }$ is to form the divergence of some current, one can show that
two terms emerge to play this role: $\vec{\nabla}\cdot \left( \rho \vec{%
\nabla}S\right) $ and $\Delta \rho $. One obtains these in a unique way by
putting $a_{1}=a_{2}=a$ and $a_{4}=a_{5}=0$. Renaming $a_{3}D\rightarrow
D^{\prime }$ and $aD\rightarrow D$ allows us to write the modified
continuity equation as\footnote{%
The term associated with $D$ does not appear in the continuity equation of
the original Doebner-Goldin modification. To restore the correspondence with
the Doebner-Goldin modification one needs to discard $d_{3}R^{2}(\vec{\nabla}%
S)^{2}$ in the Lagrangian (49), which will lead to even a more restricted
version of this modification.} 
\begin{equation}
\frac{\partial \rho }{\partial t}+\frac{\hbar }{m}\vec{\nabla}\cdot \left(
\rho \vec{\nabla}S\right) +D\vec{\nabla}\cdot \left( \rho \vec{\nabla}%
S\right) +D^{\prime }\Delta \rho =0.  \label{47}
\end{equation}
It is now straightforward to check that the following Lagrangian density 
\begin{eqnarray}
L_{DG}^{r}(\Psi ,\Psi ^{*}) &=&L_{SE}(\Psi ,\Psi ^{*})+c_{1}\Psi ^{*}\Psi
\Delta \ln \frac{\Psi ^{*}}{\Psi }+c_{2}\Psi ^{*}\Psi \Delta \ln a^{\prime
}\Psi ^{*}\Psi +c_{3}\Psi ^{*}\Psi (\nabla \ln b^{\prime }\Psi ^{*}\Psi
)^{2}+  \nonumber \\
&&c_{4}\Psi ^{*}\Psi (\vec{\nabla}\ln b^{\prime }\Psi ^{*}\Psi )\cdot \vec{%
\nabla}\ln \frac{\Psi ^{*}}{\Psi }+c_{5}\Psi ^{*}\Psi \left( \vec{\nabla}\ln 
\frac{\Psi ^{*}}{\Psi }\right) ^{2}  \label{48}
\end{eqnarray}
leads to a restricted version of the DG modification characterized by $%
b_{2}=2b_{4}+b_{3}=0$. This becomes clear when $L_{DG}^{r}$ is put in the
hydrodynamic form 
\begin{equation}
L_{DG}^{r}(R,S)=-L_{LSE}(R,S)+d_{1}R^{2}\Delta S+d_{2}\left( \vec{\nabla}%
R\right) ^{2}+d_{3}R^{2}(\vec{\nabla}S)^{2}  \label{49}
\end{equation}
that derives from the previous expression for $L_{DG}^{r}$ after dropping
total derivatives like $\vec{\nabla}\cdot (R^{2}\vec{\nabla}S)$ and $\Delta
R^{2}$, and noting that $2\Delta R$, which one obtains from $\left( \vec{%
\nabla}R\right) ^{2}$, is the same as $\left[ \Delta \rho /\rho -\left( \vec{%
\nabla}\rho /\rho \right) ^{2}/2\right] R$. Here $c_{i}$'s and $d_{i}$'s are
arbitrary dimensional coupling constants, complex (in general) and real,
respectively. The Lagrangian formulation of the entire Doebner-Goldin
modification has not been found yet, and it is conceivable that it does not
exist. The Lagrangian for the restricted variant in question, (47) or (48),
has not been presented in the literature before. We also note that the
nonlinear part of Lagrangian (48) represents a special case of the general
nonlinear Lagrangian for which the field-theoretical energy functional is
identical to its quantum-mechanical one \cite{Pusz4}.

As is well known, the Doebner-Goldin modification is both weakly separable
and homogeneous of degree one in $\Psi $, in which it differs from other
special versions of our modification for $n\geq 2$. The straightforward
multi-particle extensions of these versions being homogeneous are not weakly
separable. This is at variance with the hope expressed by Weinberg that the
homogeneity of nonlinear modifications of the Schr\"{o}dinger equation, that
plays a crucial role in his scheme of nonlinear quantum mechanics \cite
{Wein1}, guarantees the weak separability of such systems. As stated in \cite
{Wein2} : ``The problem of dealing with separated systems has led other
authors to limit possible nonlinear terms in the Schr\"{o}dinger equation to
a logarithmic form;'' and ``The homogeneity assumption (2) makes this
unnecessary.'' Even if this holds true in the Weinberg modification of
nonlinear quantum mechanics, the example of the discussed versions of our
modification shows that this is not always the case. In general, the
separability is independent of homogeneity. This was explicitly demonstrated
in the previous section for a particular case of the modification
originating from $L_{2}$. However, as we also demonstrated it in there, it
is possible to construct a unique weakly separable multi-particle extension
of these versions.

Out of the modifications mentioned, only the Bia\l ynicki-Birula and
Mycielski and Weinberg ones have been confronted with some sort of
experimental data and thereby upper bounds on their parameters have been
imposed. It seems that the former, to quote \cite{Bial2}, ``has been
practically ruled out by extremely accurate measurements of neutron
diffraction on an edge.'' The measurements in question \cite{Gahl} (see also 
\cite{Klein, Zeil} for a more comprehensive discussion of these and other
relevant techniques and experiments) have established an upper limit on the
only nonlinear parameter of the modification to be $3.3\times 10^{-15}$ eV.
The Weinberg general scheme leads to a number of particular models of
nonlinear Schr\"{o}dinger equation. Several experiments \cite{Boll, Chup,
Wal, Maj}, based on different ideas and using different techniques, were
carried out to test the most basic of these models in some simple physical
situations. For instance, in one of them, the transitions between two atomic
levels were examined to see if the resonant absorption frequencies of these
transitions were affected by the nonlinearity of the models \cite{Wein2,
Wein1}. The experiments implied an upper limit for the nonlinearity
parameter of the models to be of the order of $|1\times 10^{-20}|$ eV \cite
{Maj, Boll} which is the most stringent upper bound on nonlinear corrections
to quantum mechanics yet to date.

The Doebner-Goldin modification has not been subjected to any experimental
verification yet. As shown in \cite{Dod1}, one can, in principle, test this
modification in some experimental setup by measuring the component of the
electric current parrallel to a uniform electric field which is crossed with
a magnetic field. This current is perpendicular to both fields when all
nonlinear parameters of the theory vanish. Some indirect evidence for the
validity of the modification in question can also be gained by extending it
as proposed in \cite{Dod2} and studying the response of a two-level system
to a monochrome excitation. The nontrivial predictions regarding the
dependence of the response function on the frequency of excitation can be
experimentally verified, and if found correct, they would put the
Doebner-Goldin modification on a stronger foothold.

As far as SMPE is concerned, it remains to be investigated whether it would
be possible to determine the upper limit for the only parameter of this
modification for which we can conveniently choose the Compton quotient, the
pure number $q$, and how this modification could manifests itself in
physical phenomena. This number can, in principle, vary from one class of
objects to another.

\section{Conclusions}

We presented a possible nonlinear modification of the Schr\"{o}dinger
equation utilizing the phase of the wave function to introduce the
nonlinearity in question. Although a variation on the original proposal of
Staruszkiewicz, it does possess a number of distinctive features that in our
opinion make it deserve a more elaborate investigation, especially as far as
the simplest minimal phase extension is concerned. Unlike Staruszkiewicz's
theoretical construction, ours is independent of the space-time
dimensionality. It modifies the original Schr\"{o}dinger equation by
admitting corrections involving both the phase and the amplitude of the wave
function. The unlimited, in principle, number of terms with dimensional
coefficients that the general scheme of the modification introduces can be
easily handled by restricting our attention to the simplest models. Only
such models ensure a reasonably small departure from the Schr\"{o}dinger
equation. This leads us to the SMPE, a phase analogue of the unique
nonlinear extension of Bia{\l }ynicki-Birula and Mycielski in the realm of
the amplitude. This variant of the modification preserves all physically
relevant features of the Schr\"{o}dinger equation, compromising only the
weak separability of composed systems in its most straightforward
multiparticle extension. It is, however, possible to construct the unique
multiparticle extension that is weakly separable. The strong separability
which, as opposed to the weak one, applies also to entangled states is
maintained in the modification in question, at least in its effective
formulation \cite{Czach1}. One could avoid this compromise by breaking the
time-reversal symmetry of the free Schr\"{o}dinger equation, which would
lead us to the Doebner-Goldin modification or higher order irreversible
modifications that emerge from Lagrangian densities $L_{k1}$. However, this
would also result in giving up stationary states of quantum-mechanical
systems well established by the standard linear theory and confirmed
experimentally. We find this alternative too limiting. Instead, the SMPE
preserves these states, suggesting in addition that there may exist
stationary states beyond those predictable by linear quantum mechanics. The
modification discussed also offers an attractive minimal way of
supplementing the quantum potential of Bohm by the term that depends on the
phase of wave function. This removes a striking asymmetry of this potential
as a functional of the amplitude only. Moreover, it too removes the
asymmetry between the continuity equation and the other equation in the
Bohm-Madelung formulation by furnishing the probability current with a
``quantum'' component of the same order in the Planck constant as the
quantum potential. In general, the SMPE does not posssess the classical
limit in the sense of the Ehrenfest theorem, which suggests that it is not
linearizable and thus can describe some new phenomena that cannot be
captured by the linear theory. We have argued that the most natural way to
interpret this modification is as the simplest model of quantum mechanics of
extended objects of mass $m$ and some characteristic size $l_{c}$ to which
the coupling constant of the modification is proportional. Other
modifications that derive from the Lagrangians of $n\geq 2$ can be ascribed
the same interpretation. An alternative more traditional interpretation of
the SMPE in terms of a universal coupling constant that does not emphasize $%
l_{c}$ is also possible. The general scheme of the proposed modification
gives rise to a large class of particular models, the SMPE being only one of
them, which demonstrate that the homogeneity of nonlinear Schr\"{o}dinger
equation is not sufficient to ensure the weak separability of composite
systems contrary to the belief expressed in the literature \cite{Wein1,
Wein2}. Yet, it is possible to formulate multiparticle extensions of these
models that are weakly separable. Last but not least, all the models of the
modification possess rather an exceptional property among nonlinear
modifications of this equation in that each model's field-theoretical and
quantum-mechanical energy functional are one and the same thing \cite{Pusz4}.

There are several goals one would like to achieve through the study of the
modification proposed, the SMPE in the first place. One is a better
understanding of possible ways in which the phase of wave function can
manifest itself in quantum-mechanical systems. Due to a pervasive role the
phase plays in quantum mechanics, it is rather hard to overestimate the
importance of this question. In fact, this aspect has attracted a
considerable attention since the discovery of the Aharonov-Bohm effect \cite
{Ahar} and the Berry phase \cite{Berr} subsequently generalized by Aharonov
and Anandan \cite{Anan1}. As shown in these papers (see also \cite{Anan2}%
\thinspace for a more comprehensive review), the phase can have a nontrivial
impact on the evolution of quantum systems. One can use the models presented
in this work as a laboratory for exploring new conceivable ways of how the
phase can affect the evolution in question.

In this context, let us note that the solutions to the models mentioned do
not contain ordinary Gaussian wave packets even though the plane wave
solution is supported by all the modifications in our general scheme. This
fact is rather generic for nonlinear generalizations of the Schr\"{o}dinger
equation. As is well known, the wave packets in standard quantum mechanics
have one disturbing property: they spread beyond limit implying that there
may exist macroscopically extended quantum objects, contrary to the
experimental evidence gathered so far. Therefore, the exclusion of these
solutions should by no means be construed as a defect of the modification,
but perhaps rather as a desirable feature of it. It is also for this reason
that the wave packets of linear quantum theory can serve only as very crude
models of particle-like configurations. The localization of wave packets,
however poor it may be, is due to the interference of elementary waves and
so the phase plays a crucial role in it. One can thus wonder if the
nonlinearity of the modified Schr\"{o}dinger equation can work in synergy
with this localizing effect of the phase. We find this rather plausible. We
believe that since the SMPE can be thought of as a quantum theory of
extended particles, it is very natural to expect that it possesses localized
solutions that would describe such particles. In fact, without such
solutions the very idea of the SMPE as a theory of extended objects appears
to be rather vacuous. Their existence would provide a viable particle
representation of the wave-particle duality embodied in quantum mechanics.%
\footnote{%
Such solutions have indeed been found \cite{Pusz3} in the form of Gaussian
solitons.} As of now, it is only the wave representation of this duo that is
supported by the current understanding of quantum theory. Although it is
premature to speculate on it, it is not entirely out of the question that
the SMPE as a theory of extended objects might be useful in the description
of some classes of such objects as, for instance, nuclei.

Another general and important motivation to study this modification comes
from the long-standing problem of the collapse of wave function. The linear
Schr\"{o}dinger equation is incomplete in the sense that it does not
describe this process. Therefore, it has been conceived in the literature
that introduction of nonlinear terms can solve this problem. Various
particular models incorporating such terms have been proposed, to name just
a few of them \cite{Pea1, Grig, Hugh1, Five, Pusz5} (for a review see \cite
{Pea2, Ghir}). Since it is certainly reasonable to expect that the phase
plays a significant role in this process, we believe that a further
exploration of the modification presented could cast more light on possible
mechanisms of this phenomenon.

As noted in \cite{Bial}, the Schr\"{o}dinger equation is one of rare
examples of fundamental equations of physics that are apparently linear. One
can however entertain the thought that this so only in some leading
approximation to a more general nonlinear equation that still awaits its
discovery. No fundamental proof to the contrary has been provided yet as
emphasized recently once more by Czachor \cite{Czach1} and Jordan \cite{Jord}%
. In fact, nonlinear extensions of this equation have found some
applications, typically in the description of collective phenomena. Of
particular use here has been the cubic nonlinear Schr\"{o}dinger equation
that naturally emerges in the mean-field approach to many-body problems in
quantum mechanics. For instance, this equation appears in the theoretical
treatment of Bose-Einstein condensation which, as emphasized in \cite{Grif},
``is a common phenomenon occuring in physics on all scales from condensed
matter to nuclear, elementary particle and astrophysics.'' In this case, the
nonlinear Schr\"{o}dinger equation is often referred to as the
Gross-Pitaevskii equation \cite{Gross1, Gross2, Pita}. It describes a
macroscopic population of particles, be it atoms, kaons or excitons in
semiconductor heterostructures, condensed in the ground state of a system at
the absolute zero temperature and interacting weakly through a short range
potential modeled by a Dirac delta term. Yet another application for this
equation has been found in nuclear physics where it was invoked to account
for the compressibility of nuclear matter in elastic scattering of heavy
ions \cite{Del}. Interestingly, this equation also tends to appear in
schemes involving gravity. In such situations \cite{Rosa, Nam, Ham}, the
nonlinear term emerges as a correction to the linear equation dictated by
the effect of the gravitational field of the particle on its own wave
function, i.e., via the backreation. Nevertheless, due to the smallness of
the Newton constant, the gravitationally induced correction is practically
negligible. As these examples demonstrate, even if the Schr\"{o}dinger
equation is never found truly fundamentally nonlinear, the
quantum-mechanical picture of reality can equally well be found incomplete
without nonlinearities invoked to account for particular physical effects,
especially that some of these effects, even though negligibly small, appear
as approximations in a grander scheme of things, as in the framework of
general relativity. It would be interesting to find whether the SMPE can be
an approximation to a more all-encompassing theoretical model.

However, from a fundamental point of view, as accentuated by Weinberg \cite
{Wein1}, nonlinear generalizations of the Schr\"{o}dinger equation can play
a useful role as a kind of foil against which to test the linearity of
quantum mechanics. Such a role has already been successfully played by the
Bia\l ynicki-Birula and Mycielski modification and even to a greater extent
by a particular model of Weinberg's proposal. It is in this context that one
can envisage another use of the modification proposed. More studies are
needed to examine whether, and if so, how the SMPE could be employed to set
up a bound on the nonlinearity parameter it introduces, similarly as it was
done in the modifications just mentioned.

Yet another major goal of studying this modification is a more profound
examination of the original phase modification of Staruszkiewicz, the
modification of Doebner and Goldin whose restricted but fairly general
version appears as a rather natural extension of a particular model of our
general scheme, and its generalization recently proposed by this author \cite
{Pusz2}. Although these modifications differ from the SMPE that we consider
the most interesting proposal presented here, one can expect that some
insight gained from the study of this modification may also be applicable to
them. Since, unlike the other modifications, the Staruszkiewicz modification
is nonhomogeneous in the wave function, one hopes that comparative studies
of these modifications could cast more light on the issue of physical
consequences of nonhomogeneity in a modified Schr\"{o}dinger equation.

Further investigations of the SMPE including in particular some physically
interesting solutions will be reported elsewhere \cite{Pusz3, Pusz6}.

\section*{Acknowledgments}

I would like to thank Professor Pawe{\l } O. Mazur for bringing to my
attention the work of Professor Staruszkiewicz that started my interest in
nonlinear modifications of the Schr\"{o}dinger equation and Professor Jeeva
Anandan for a discussion of the importance of phase in quantum mechanics. I
am particularly grateful to Professor Andrzej Staruszkiewicz for a critical
reading of the preliminary version of this paper, his comments and a
discussion. A stimulating exchange of correspondence with Dr. Marek Czachor
and a correspondence with Professor Wolfgang L\"{u}cke regarding his recent
work are also greatfully acknowledged. This work was partially supported by
the NSF grant No. 13020 F167 and the ONR grant R\&T No. 3124141.

\end{document}